\def\year{2020}\relax
\documentclass[letterpaper]{article} 
\usepackage{aaai20}  
\usepackage{times}  
\usepackage{helvet} 
\usepackage{courier}  
\usepackage[hyphens]{url}  
\usepackage{graphicx} 
\usepackage{graphicx}  
\title{War of the Hashtags: Trending New \\ Hashtags to Override Critical Topics in Social Media }
\author{Debashmita Poddar\\ 
Gran Sasso Science Institute\\ 
L'Aquila, Italy\\
debashmita.poddar@gssi.it 
}
 
\begin{document}

\maketitle

\begin{abstract}

\emph{Hashtags} play a cardinal role in the classification of topics over social media. A sudden burst on the usage of certain hashtags, representing specific topics, give rise to \emph{trending topics}. Trending topics can be immensely useful as it can spark a discussion on a particular subject. However, it can also be used to suppress an ongoing pivotal matter. 
This paper discusses how a significant economic crisis was covered by triggering a current trending topic. A case study on politics in India has been studied over the past two months. The analysis shows how the issue on inflation was attacked by the exercise of a new constitutional law over media. Hashtags used to discuss the topics were scrutinized, and we notice a steep ascend of the more recent topic and an eventual drop in discussions over the previous issue on inflation. 
Balancing the influence of hashtags on social media can be employed. Still, it can be equally challenging since some hashtags that represent the need of the hour topics should be given more importance, and evaluating such issues can be hard. 

\end{abstract}

\section{Introduction and Motivation}

Social networking sites have seen a notable boom in the past few years. Indeed, more and more people are taking part in the services provided by these sites. Some of the major networking sites with over a hundred million active users include \emph{Facebook}, \emph{Twitter}, \emph{Instagram}, etc. A social network is a collection of entities which form relationships with other entities participating in that network. \emph{Social Network Analysis (SNA)} has emerged due to the increase of information propagation in the networking sites. SNA is the study of social networks using collaboration graphs in which nodes of a graph model the entities, and edges models the relationships between entities.

Social media plays a crucial role in spreading new ideas or innovations through its fundamental network structure consisting of individuals and their relationship with other individuals. 
Henceforth, most companies and politicians have resorted to social media for marketing products and advertising political ideologies, respectively. However, the core of these two scenarios lies in reaching the maximum number of users in a given network. 
In marketing, the company has a specific budget to sponsor a handful number of people, also called the \emph{influencers}, who has a certain degree of influence on the other participating entities in the network. The goal is to influence the maximum number of entities of this network to adopt the marketed product. 
In the case of \emph{political campaigns}, the politicians need to spread their ideology to a targeted audience who can be potential voters for the politician. The idea is to select a batch of people who associates with the political party’s ideologies and volunteers to promote the party among the general masses. Since the volunteers are influencers, the goal sticks to influence as many people as possible in the network.

Some of the leading social media like Twitter popularised the hashtag trend. A hashtag is a collection of characters that can sum up to a word or a phrase preceded by a number sign (\#). It acts as a tag to group related contents. Hashtags are generally used for promotions; it can be an event, a campaign, complaints, or a specific product. For example, the hashtag, \emph{\#icwsm19} will show all the contents related to the \emph{13th ICWSM-2019} held in Munich, Germany that were tagged with this specific hashtag. Netizens can voice their opinions or ideas and can follow it up with a conditioned hashtag. It also plays a crucial part in the analysis of social media due to its data clustering property. 

Political campaigns depend heavily on hashtags. For example, in the \emph{2016 U.S elections}, along with \emph{\#Election2016}, the two political parties used the \emph{\#VoteTrump} and \emph{\#ivotedforHillaryClinton} respectively to reach out to the general public on social media. 
In Twitter, for a hashtag to qualify as a trending topic, the Twitter trending algorithm is designed in such a way that the users will be shown the trending tweets based on their location, interests and the profiles they follow. \emph{Trends} are usually the hashtags that have been used vastly in a short period. Facebook also has a trending module to show what topics are being discussed. Trends are an adequate way to ignite a discussion on a topic.

However, considering the dynamic and erratic nature of social media, a trending topic can get replaced within seconds of its initiation; this can have a disadvantage over a current problem, which is trending. If the issue at hand carries a significant concern and needs to reach out to its concerned audience, another trending topic can be instigated to reduce the impact of the former on the audience.
Previous research has shown the fleeting nature of trending topics. Extensive studies on Twitter has been conducted to give quantitative insights on trending topics \cite{Kwak2010}. Surveys on the temporal dynamics of tweets have been carried out by analysing the tweets in real-time search system to learn more about how the trending topic mechanisms work \cite{Lin2012}. Inspection of the lag between the inception of trending topics and the hunt for those trending topics is also an area of concern \cite{Kairam2013}. The delay could be useful to understand the dynamics between why some tweets are better received by individuals and have a higher chance of being on the trending list.

In this paper, I will introduce a case study on \emph{Indian Politics}, that will help us understand the pre-eminent nature of hashtags, and how it can affect a political scenario of a country. The paper will also throw some light upon how powerful hashtags can be when it is used strategically, and the way it affects the everyday lives of people outside social media. 

Two open-ended research questions follow immediately after the discussion:\\
\textbf{RQ1}- Is there a way to limit the influence of newer trending hashtags, that have the potential to overpower the current topic in hand?\\
\textbf{RQ2}- If so, how do we determine which hashtags come as a disguise to suppress the contemporary hashtags?

This paper introduces some concepts of \emph{influence maximization} by a potential set of nodes \cite{Kempe2015}, and how these influential nodes can be used to bring balance to the information propagation in the network \cite{Garimella2017}. The studies on influence maximization can be extended to balance the effect of competing hashtags in the network. RQ1 can be seen as a balancing problem and can use the same mechanics. Nevertheless, we will also see that RQ2 can be exigent, and addressing RQ1 becomes quite pressing under RQ2. The paper lacks solutions but provides a summary of the competitive nature of hashtags to find a spot on the trending topic list.

\section{Case study: Politics in India}

Politics in India is autonomous. The citizens of India decides which government should be in power. The \emph{2019 Indian general election} saw the previous government coming into power. The \emph{Bharatiya Janata Party (BJP)} and the \emph{National Democratic Alliance (NDA)}, which is a coalition of regional state parties led by BJP and sharing the same ideologies of the BJP, won the elections by forming a coalition government. \emph{Narendra Modi} was re-elected as the Prime Minister.

A major event that struck India from October last year was the \emph{onion crisis}. Due to the delay of monsoon, it took a remarkable toll on the farmers. Not only did the unseasoned rains damage the crops, but it also reduced the supplies of onions. The price of onion was approx \emph{$200$ INR per kilogram  ($2.82$ USD approx)} as compared to the original price of approx \emph{$15$ INR per kilogram ($0.21$ USD approx)}. As a natural reaction to overcome the onion crisis, the government decided to ban the export of onions until the situation normalises; this has a direct effect on the farmers since it reduces their income from exports. Also, due to the reduction in onion production, the farmers can not sell the estimated amount of onion to the market retailers, which cuts down the farm income. The retailers, due to scarcity of onions and its high market demand, hiked the price of onions, causing dismay among customers. Due to a sudden price surge, people have to cut down on the consumption of onions, which is one of the staple vegetables in India. The farmers are incurring monetary losses due to the ban on onion trade and low farm income from the retailers and traders.

\begin{figure}[t]
\centering
\includegraphics[width=0.9\columnwidth]{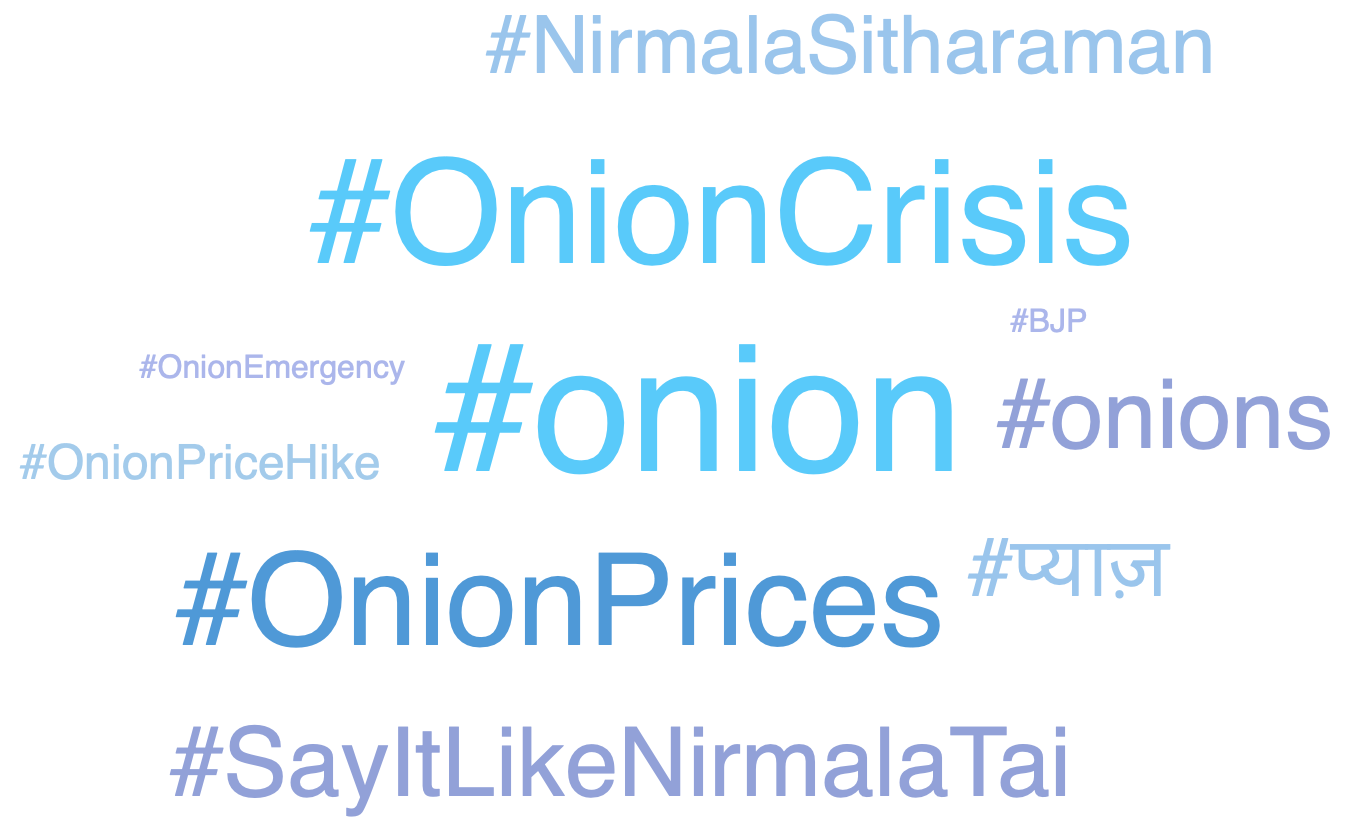} 
\caption{Other related hashtags concerning the onion price hike in India. \emph{Nirmala Sitharaman} is the Indian Minister of Finance and Corporate Affairs}
\label{fig1}
\end{figure}

In the 2019 political campaign, one of BJP’s agenda included a raise of income for the farmers and keeping a check on food inflation. As farmers form a majority in the voting constituency, Modi promised that his topmost priority would be the farmers. The onion crisis sparked rage among the citizens — consequential protests on the incompetence of the government in keeping their promises followed. Social media was flooded with disappointment. One of the trending hashtags was \emph{\#onionprice}. Figure \ref{fig1} shows the other hashtags that were used to raise the concern.

Another significant event that took place in December last year was the \emph{Citizenship Amendment Act (CAA)} which incurred a nationwide protest across India. Primarily, the \emph{Citizenship Amendment Bill (CAB)} was introduced by \emph{Amit Shah}, Minister of Home Affairs in House of the people, called \emph{Lok Sabha}, few days before enacting it. According to the CAA, any individual who migrated to India before 2015 and belonged to the specific minority category of Hinduism, Sikhism, Buddhism, Jainism, Zoroastrianism and Christianity from Afghanistan, Bangladesh and Pakistan would be granted an Indian Citizenship. The CAA claims to remove illegal immigrants but is considered to have a communal incentive attached to it since there is no mention of Muslim refugees or Tamil Hindu refugees from Sri Lanka.

The \emph{National Register of Citizens (NRC)} contains the data of all legal citizens generated by \emph{The Citizenship Act, 1955}. The Citizenship Act, 1955, states that individuals who were residents of India at the commencement of the Constitution of India are granted the status of Indian citizens. It also allows individuals to claim Indian citizenship if they are born in India. The CAA amends to the Citizenship act. Individuals not recorded in the NRC can request to have their names on it by applying the CAA. However, Muslim immigrants and Tamil immigrants can not apply since the CAA has no specific mentions about them. The constitution of India, although, mentions that India is a secular and democratic country, and each religion must be treated equally. The discrepancy caused by the CAA saw a series of ongoing protests in India. 

\begin{figure}[t]
\centering
\includegraphics[width=0.9\columnwidth]{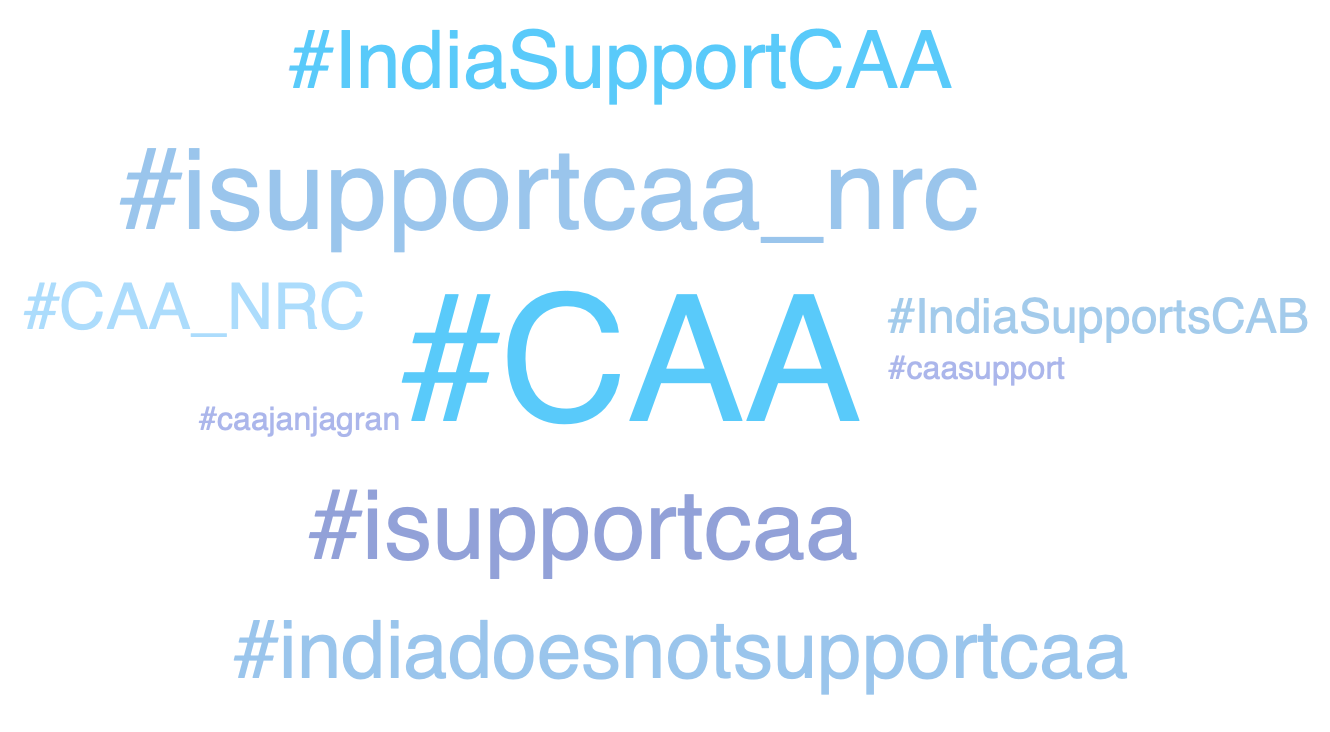} 
\caption{Related hashtags supporting the CAA}
\label{fig2}
\end{figure}

\begin{figure}[t]
\centering
\includegraphics[width=0.9\columnwidth]{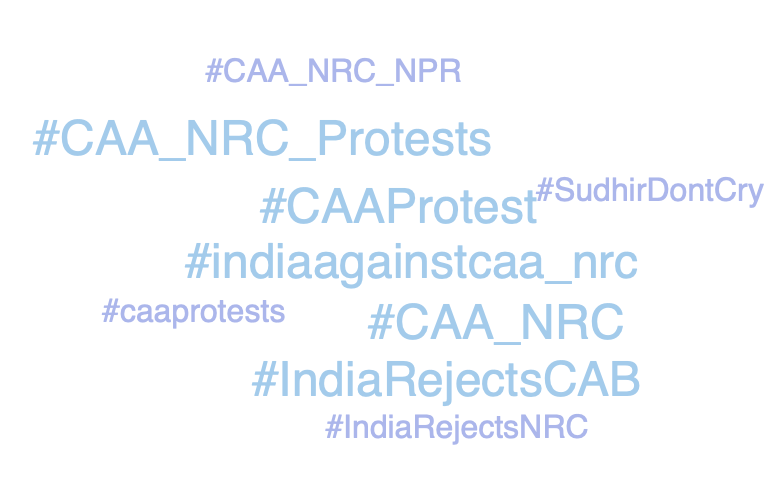} 
\caption{Trending hashtags against the CAA}
\label{fig3}
\end{figure}

The BJP government took to their social media handles, especially Twitter, to talk about the benefits of implementing CAA. The hashtag used by the BJP government to trigger the discussions on CAA was \emph{\#IndiaSupportsCAA}. Netizens immediately took the social media by storm and two parties were formed. Individuals that supported the CAA carried on with the discussions by using the \#IndiaSupportsCAA. A strong opposition developed against the CAA, and opinions were shared by trending the hashtag \emph{\#IndiaAgainstCAA} to resist the government. Figures \ref{fig2} and \ref{fig3} show the different hashtags used by people for and against the CAA.

The protests took a different direction as the focus shifted from the price hike of onion to the CAA. The topic of the onion price surge died down ultimately. From trending to dormant, we see how the use of stronger hashtags can overpower a crisis. From media houses to everyday discussions, the onion price hike topic was obliterated. The following section will detail more on how \#IndiaSupportsCAA and \#IndiaAgainstCAA overrode \#onionprice.

\section{Inferences and Interpretations}

\begin{figure}[t]
\centering
\includegraphics[width=0.45\textwidth]{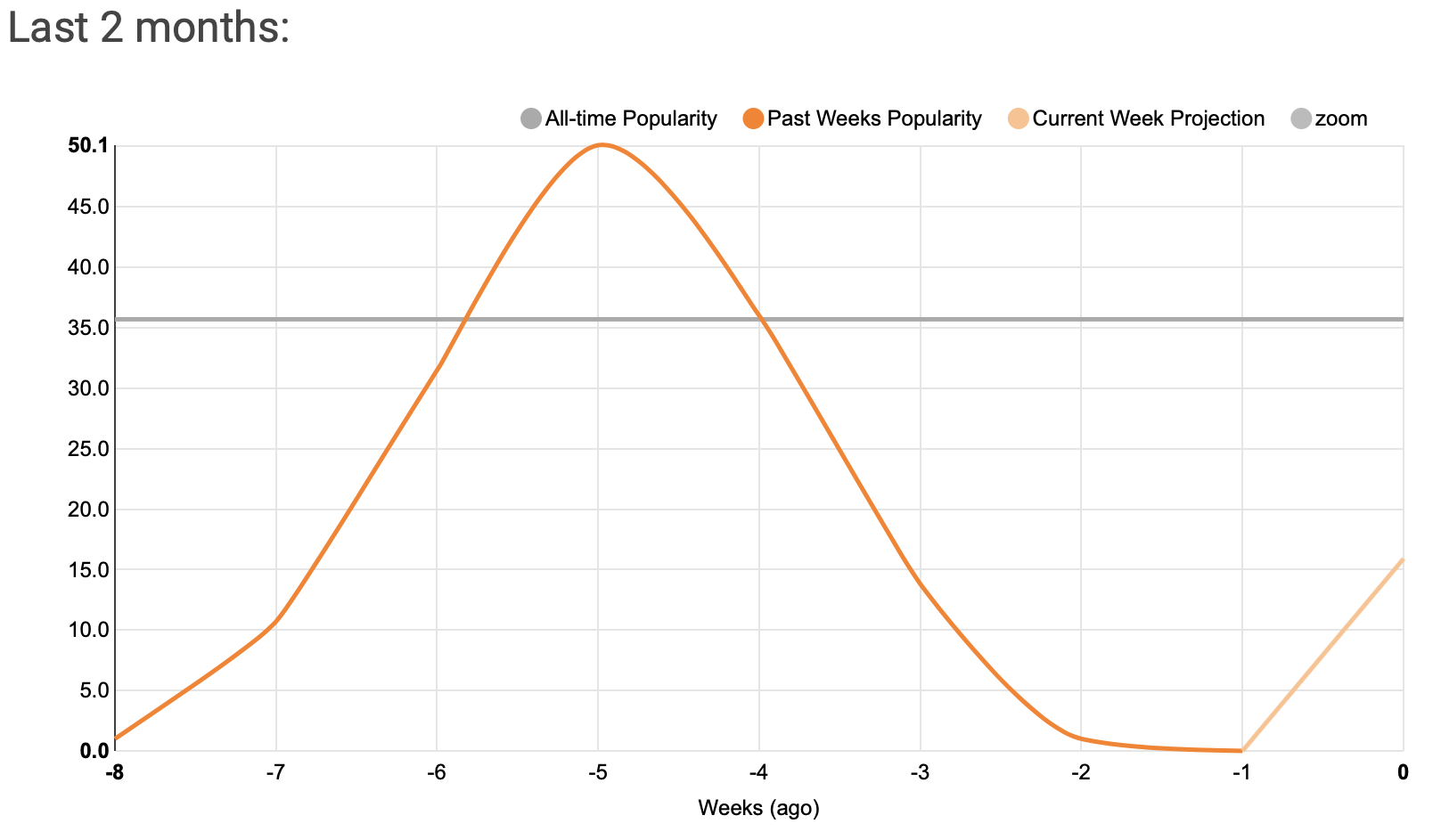} 
\caption{Graph showing the rate of mentions of \#onionprice over the last two months.}
\label{fig4}
\end{figure}

\begin{figure}[t]
\centering
\includegraphics[width=0.45\textwidth]{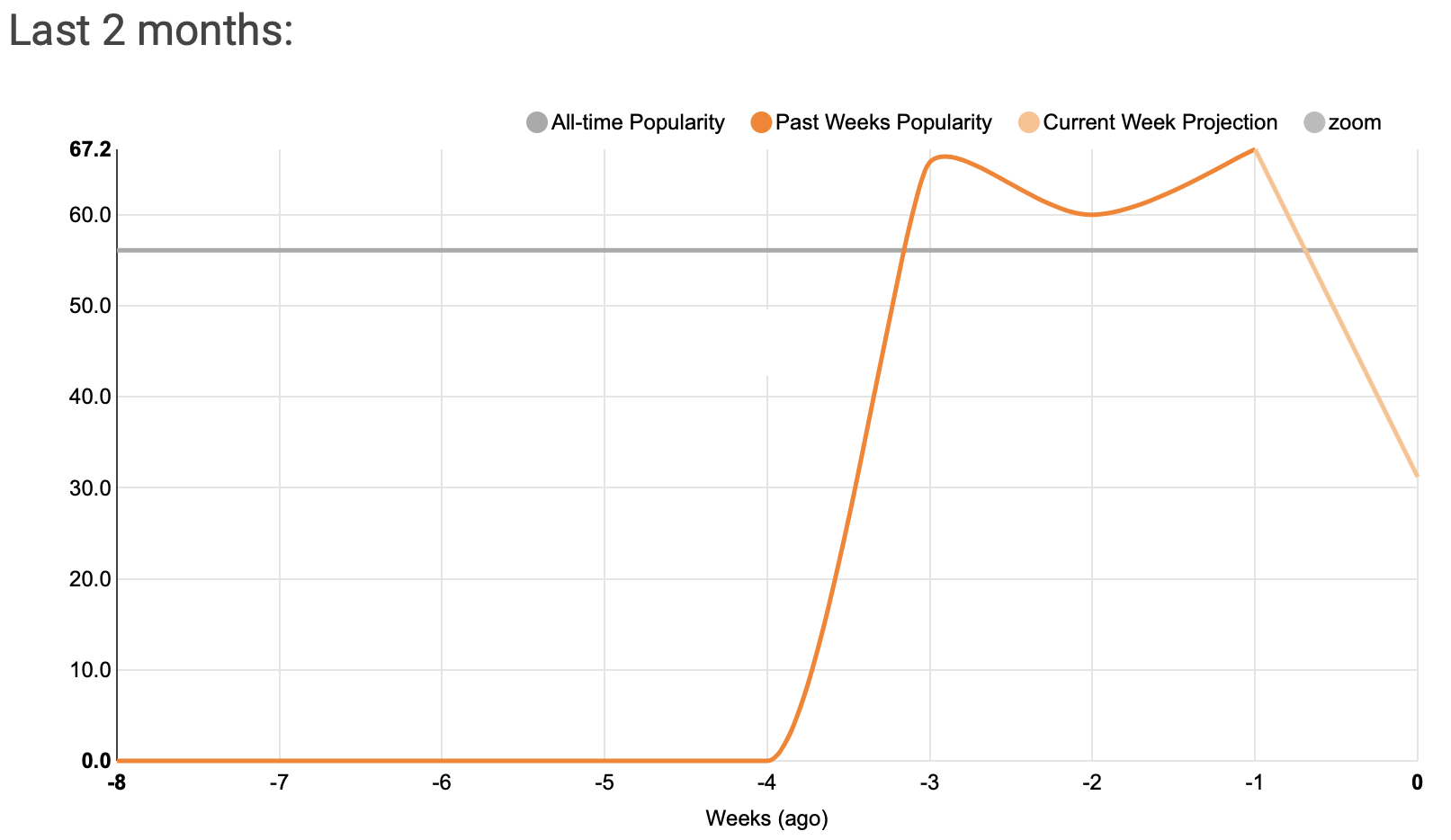} 
\caption{Graph showing the steep rise of \#IndiaSupportsCAA mentions from December 2019.}
\label{fig5}
\end{figure}

\begin{figure}[t]
\centering
\includegraphics[width=0.45\textwidth]{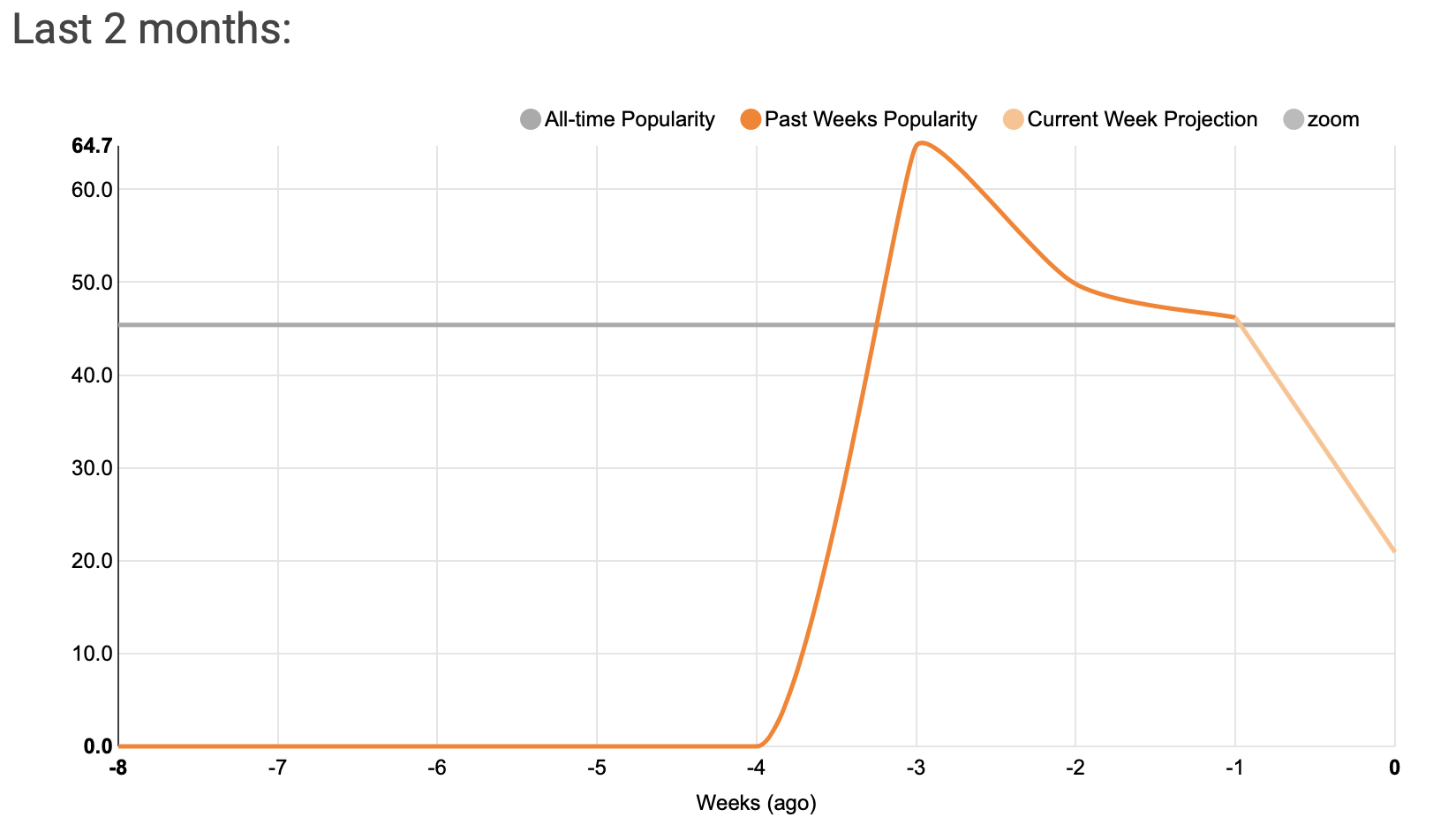} 
\caption{\#IndiaAgainstCAA also saw a steep descent alongside \#IndiaSupportsCAA.}
\label{fig6}
\end{figure}

I used \emph{Hashtagify}, which is a free online hashtag tracking tool, to analyse the three hashtags over two months (11th November 2019 -11th January 2020). Figure \ref{fig4} shows the graph of \#onionprice mentions. Figure \ref{fig5} and \ref{fig6} shows the usage of \#IndiaSupportsCAA and \#IndiaAgainstCAA. From Figure \ref{fig4}, we observe that the \#onionprice has a gradual growth from the 11th November, 2019 and continues to form an ascending curve for about another four weeks (marked as -5), then gradually descends from the start of 4th week onwards. On observing Figures \ref{fig5} and \ref{fig6}, we see that the \#IndiaSupportsCAA and \#IndiaAgainstCAA form a steep curve from the 5th week (marked as -4) and continues to linger around the higher values.  

If we take a closer look at the graphs, we see that the gradual descent of \#onionprice has been caused by the abrupt increase of discussions on \#IndiaSupportsCAA and \#IndiaAgainstCAA. Comparing the three hashtags, we see that the latter ones about the CAA strangle the former hashtag about onions. The trend was observed not only in the virtual world but also in the real world. Leading newspapers and media houses started covering news about CAA, and there was little to no mention about the onion price surge - this impacted the general public more, and the hub of agitation diverted from the economic crisis to CAA.

\subsubsection{Addressing RQ1}
The \emph{word of mouth} effect in the online marketing world \cite{Domingos2001} introduces the influence propagation on social media. Later the influence propagation was designed as a discrete optimization problem and was named as the influence maximization problem \cite{Kempe2015}. The influence maximization problem states that, given a network, we need to find a set of k nodes that have the maximum influence over the concerned network. The authors introduced some underlying diffusion models to study the influence spread of the network and proved that the maximization problem is NP-hard. The discrete optimization problem can be used to minimize or limit the effect of newer hashtags that have the potential to be the trending topics.

The \emph{Eventual Influential Limitation (EIL)} problem proposes a limiting campaign that would alleviate the effects of the bad campaign \cite{Budak2011}. Following the same path, studies have been conducted on how to balance the information exposure in social networks, to curb social homophily \cite{Garimella2017}. The general framework of balancing information exposure is considered: two sets of nodes are found, which can represent a post on each of the hashtags so that the influence made by the hashtags on individuals participating in the network is balanced. But it may be possible that some posts can use both the competing hashtags, which makes the evaluation difficult. One might use \emph{sentiment analysis} to detect the tone of the text, whether the individual has a positive, negative, or neutral attitude towards the concerned topics.

However, limiting the trending topics can be tricky. \emph{RQ2} opposes \emph{RQ1} by stating that even if we find a way to limit the influence of a trending topic, how can we determine which topic needs more focus? In the case study presented in the paper, both \#onionprice and \#IndiaSupportsCAA are essential topics. Since \#onionprice was a hashtag introduced one month before \#IndiaSupportsCAA, it’ll be difficult for the former to prolong its position in the trending topic list. It becomes onerous to find out whether a trending hashtag is conditioned to subdue the influence of a previously trending hashtag. It is quite probable that a newly trending hashtag might need the attention of the hour and its influence needs to be maximized over the network. Henceforth, RQ2 remains as a convoluted argument.

\section{Conclusion}
Social Media has a massive impact on our day to day lives. It is an essential platform for the spread of information. The usage of hashtags to classify specific topics is considered to be revolutionary. To get news on a particular subject, one can select the hashtags associated with that subject and can acquire the contents related to it. Trending topics are those hashtags that have been popularised over a short period of time. If a hashtag is in the trending topic list, it means that the matter related to the hashtag has influenced a significant amount of people over a brief period.
Furthermore, a newly trending hashtag has the power to dampen the influence of a previously trending hashtag- this can have significant implications on social media. If a topic is crucial, its importance can decrease by introducing potentially trending hashtags. 

This paper considers a particular case study on Indian Politics. It shows how a hashtag \#onionprice, which talks about the economic crisis caused in India due to the surge of onion prices, is stifled by two other hashtags, namely \#IndiaSupportsCAA and \#IndiaAgainstCAA. The latter hashtags immediately divert the attention from inflation caused due to onion price hike to a controversial constitutional act. Furthermore, we see that limiting the influence of a new trending hashtag can be intricate since it can not be assured which of the two topics is more salient.

\bibliographystyle{aaai}
\bibliography{ref}

\end{document}